\documentclass[a4paper]{jpconf}
\usepackage{graphicx}

\newcommand {\snn}      {\sqrt{s_{\rm NN}}}
\newcommand {\raa}      {R_{\rm AA}}
\newcommand {\iaa}      {I_{\rm AA}}

\newcommand {\pT}        {p_{\rm T}}

\newcommand {\dphi}     {\Delta\phi}

\begin{document}

\title{Jet-like correlations of heavy-flavor particles -- from RHIC to LHC}

\author{Andr\'e Mischke}

\address{ERC-Research Group {\em QGP-ALICE}, Institute for Subatomic Physics, Utrecht University, Princetonplein 5, 3584 CC Utrecht, the Netherlands}

\ead{a.mischke@uu.nl}

\begin{abstract}
Measurements at the Relativistic Heavy Ion Collider (RHIC) at Brookhaven National Laboratory have revealed strong modification of the jet structure in high-energy heavy-ion collisions, which can be attributed to the interaction of hard scattered partons with the hot and dense QCD matter.
The study of heavy-quark (charm and bottom) production in such collisions provides key tests of parton energy-loss models and, thus, yields profound insight into the properties of the produced matter.
The high-$\pT$ yield of heavy-flavor decay electrons exhibits an unexpected large suppression.
Since those single electrons have contributions from charm and bottom decays an experimental method is needed to investigate them separately.
Heavy-flavor particle correlations provide information about the underlying production mechanism.
In this contribution, a review on recent measurements on azimuthal correlations of single electrons and open charmed mesons at RHIC and perspectives of such measurements at the CERN-Large Hadron Collider (LHC) are presented.
Moreover, it has been shown that next-to-leading-order (NLO) QCD processes, such as gluon splitting, become important at LHC energies. It will be demonstrated how this contribution can be determined through the measurement of the {\em charm content in jets}. 
\end{abstract}

\section{Introduction} 
The energy loss of partons is predicted to be a sensitive probe of the QCD matter created in high energy nucleus-nucleus collisions since its magnitude depends strongly on the color charge density of the matter traversed~\cite{mueller06, jacobs05}.
In particular, the understanding of the flavor dependent coupling of quarks and the modification of their fragmentation function give essential information on the properties of the hot QCD matter produced in such collisions~\cite{ov:rapp, ov:frawley}. 
Due to their large mass ($m >1.2$ GeV/$c^2$), heavy quarks are believed to be produced predominantly in hard scattering processes in the early stage of the collision and, therefore, probe the complete space-time evolution of the expanding medium and their yields are sensitive to the initial gluon density~\cite{lin95}. 
Heavy-quark production by initial state gluon fusion also dominates in heavy-ion collisions where many, in part overlapping nucleon-nucleon collisions occur.
It has been shown that charm production in the QCD medium might be significant at LHC energies~\cite{theo:uphoff}.
The penetrating power of heavy quarks is much higher than for light quarks, providing a sensitive probe of the matter.
Theoretical models based on perturbative QCD predict that the energy loss of heavy quarks in the medium is expected to be smaller compared to light-quarks and gluons due to the mass dependent suppression of the gluon radiation at small angles, the so-called dead-cone effect~\cite{deadcone1, deadcone2}.

However, recent RHIC results on single electron production in nucleus-nucleus collisions have shown that the yield at high transverse momentum is suppressed to the same level as observed for light-quark hadrons.
Azimuthal angular correlations of heavy-quark particles allow for the study of the fragmentation function of charm and bottom quarks separately, where high-$\pT$ decay electrons are associated with open charmed mesons.

\section{Single electron $\raa$} 
\begin{figure}[t!]
\centering
\includegraphics[scale=0.38]{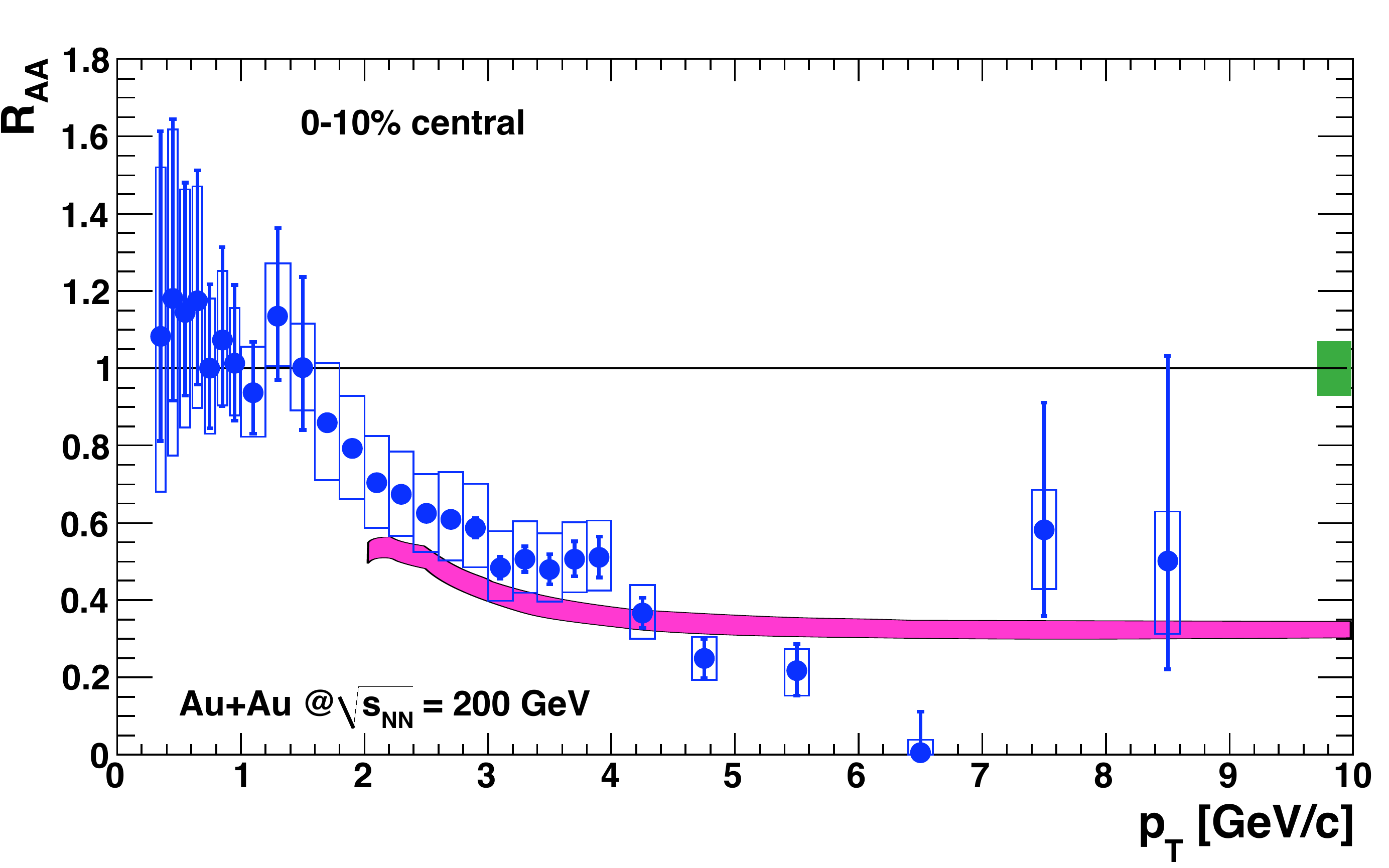}
  \caption{Nuclear modification factor $R_{\rm AA}$ of single electrons in the 10\% most central Au+Au collisions at $\sqrt{s_{\rm NN}}$~= 200 GeV measured by the PHENIX Collaboration. The data are compared with a collisional dissociation model (purple curve).} \label{Fig:1}
\end{figure}
At RHIC charm and bottom quarks are identified by assuming that isolated electrons in the event originate from semileptonic decays of heavy-quark mesons. 
At high transverse momentum, this mechanism of electron production is dominant enough to reliably subtract other sources of electrons such as $\gamma$ conversions and $\pi^0$ Dalitz decays.
STAR~\cite{Star:npe} and PHENIX measurements~\cite{Phe:npe} in central Au+Au collisions have shown that the high-$\pT$ yield of electrons from semileptonic charm and bottom decays is suppressed relative to properly scaled proton-proton collisions, quantified in the nuclear modification factor $\raa$~\cite{Star:npe} (cf. Figure~\ref{Fig:1}). 
This factor exhibits an unexpectedly similar amount of suppression as observed for light-quark hadrons, suggesting substantial energy loss of heavy quarks in the produced medium.
This finding is not in line with expectations from the dead-cone effect.
Energy-loss models incorporating contributions from charm and bottom do not describe the observed suppression sufficiently well (see discussions in~\cite{Star:npe, Phe:npe}). 
Although it has been realized that energy loss of heavy quarks by elastic parton scattering causing collisional energy loss is probably of comparable importance to energy loss by gluon radiation, yet the quantitative description of the suppression is still not satisfying.
First attempts has been performed to distinguish between radiative and collisional energy loss~\cite{theo:aichelin}.
Furthermore, it has been shown that the in-medium fragmentation of heavy quarks into heavy mesons which are collisionally dissociated, a process that can happen multiple times in the medium, may be significant in heavy-ion collisions.
In addition, cold nuclear effects might explain part of the observed high-$\pT$ yield suppression~\cite{durham}.

\section{Azimuthal correlations of heavy-flavor particle}
\begin{figure}[t]
\centering
   \includegraphics[scale=0.41]{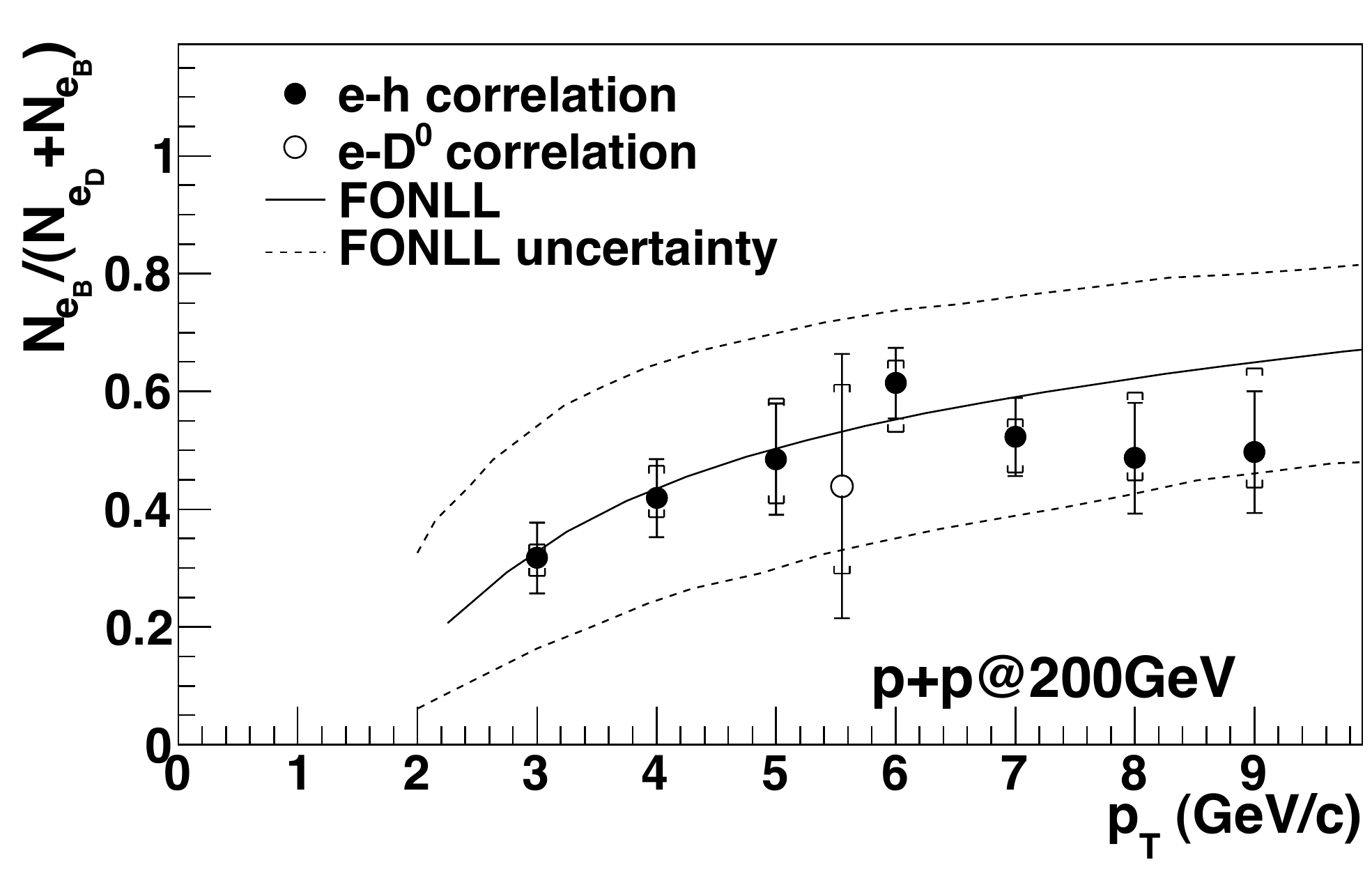}
   \caption{Relative bottom contribution to the single electron yields as a function of transverse momentum in $pp$ collisions at $\sqrt{s}$~= 200 GeV obtained from electron--hadron and electron--D$^0$ meson azimuthal correlations. Error bars are statistical and brackets are systematic uncertainties. The solid curve illustrates the FONLL calculation~\cite{theo:Mat05} with its theoretical uncertainties (dashed curve).}  \label{Fig:2a}
\end{figure}
Perturbative QCD calculations at Fixed-Order plus Next-to-Leading Logarithm (FONLL) level have shown that the bottom contribution to the heavy-quark decay electrons is significant at intermediate $\pT$ (4-5 GeV/c) although the uncertainties are relatively large~\cite{theo:Mat05}.
Therefore, measurements of the relative charm and bottom contribution to the single electrons is essential. 
Azimuthal angular correlations of single electrons and hadrons allow to exploit the different fragmentation of the associated jets~\cite{sickles}. Heavy quarks have, in general, a harder fragmentation function than gluons and light quarks, making the near-side correlation more sensitive to the decay kinematics. For the same electron transverse momentum the near-side electron--hadron angular correlation distribution from B decays is much broader than that from D decays.

Furthermore, azimuthal angular correlations of heavy-quark decay electrons and open charmed mesons yield important information about the charm and bottom contributions due to their different decay kinematics~\cite{Misch09}.
Flavor conservation implies that heavy quarks are produced in quark anti-quark pairs. A more detailed understanding of the underlying production process can be obtained from events in which both heavy-quark particles are detected. Due to momentum conservation, these heavy-quark pairs are correlated in relative azimuth ($\dphi$) in the plane perpendicular to the colliding beams, leading to the characteristic back-to-back orientated sprays of particles. This correlation survives the fragmentation process to a large extent in $pp$ collisions.
Charm quarks predominantly ($\approx$54\%) hadronize directly and bottom quarks via B decays into D$^0$ mesons. The branching fraction for charm and bottom decays into electrons is $\approx$10\%~\cite{pdg2010}. While triggering on the so-called leading electron (trigger side), the balancing heavy quark, which is identified by the D$^0$ meson, can be used to identify the underlying production mechanisms (probe side).
Requiring like-sign electron--kaon pairs, where the kaons stem from the D$^0$ decays, the near-side correlation peak can be attributed to B decays only whereas the contribution to the away-side peak can mostly be attributed to charmed meson decays.
Figure~\ref{Fig:2a} depicts the relative bottom contribution to the single electrons from $pp$ collisions at $\sqrt{s}$~= 200~GeV measured by the STAR experiment~\cite{Star:ehpaper}. The PHENIX Collaboration obtained similar results~\cite{Phe:ehpaper}.
The data provide convincing evidence that the bottom contribution to the single electron yields is of the same magnitude than that of charm at and above $\pT$ = 5 GeV/c.
The direct measurements of open charm and bottom mesons at RHIC will be possible with the high precision $\mu$ vertex detectors in the STAR and PHENIX experiments.
\begin{figure}[t]
\begin{center}
   \includegraphics[scale=0.44]{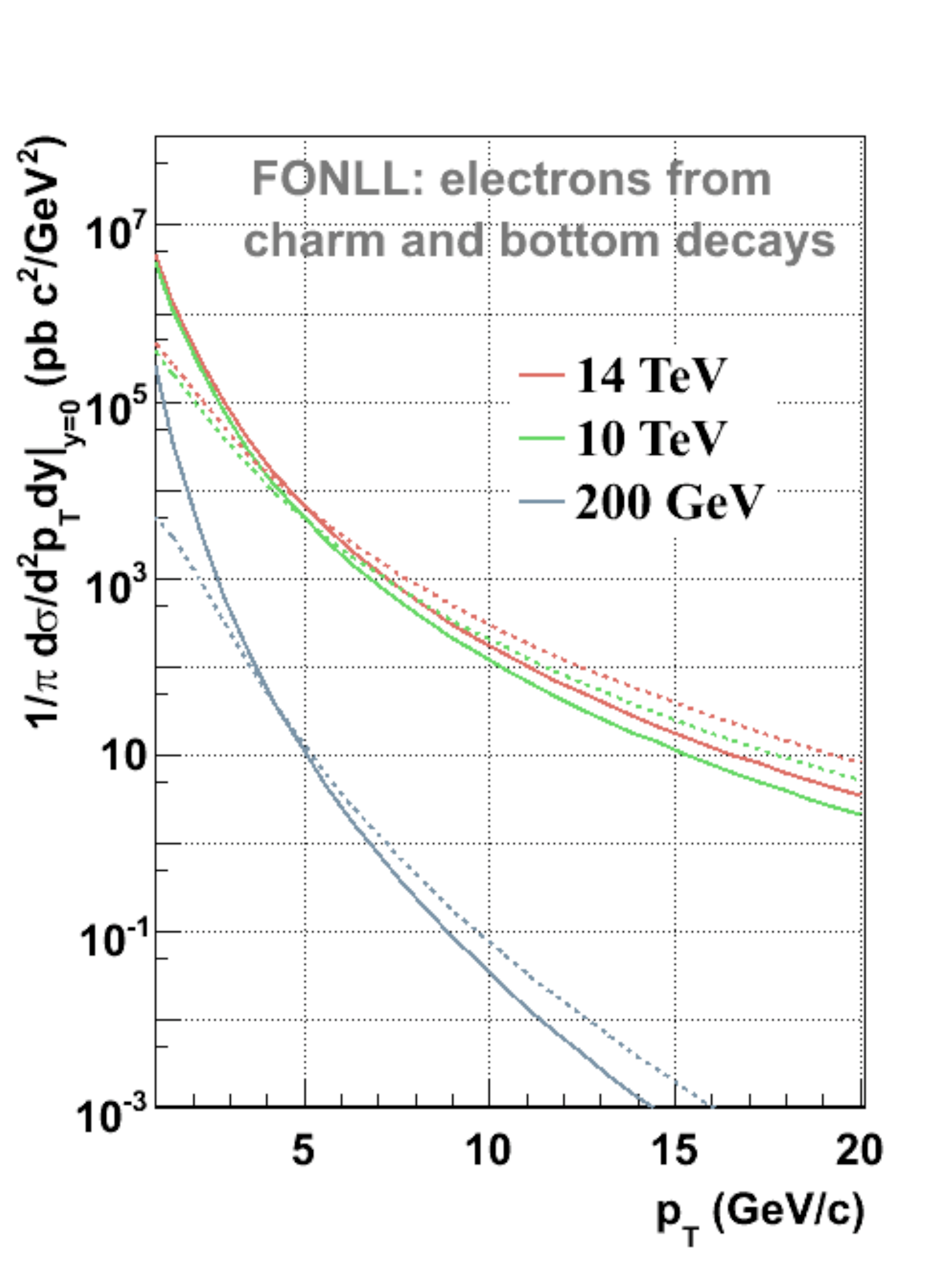}
   \begin{minipage}[b]{14pc}\caption{Electron production cross section from charm (full line) and bottom decays (dashed line) in $pp$ collisions at RHIC and LHC energies obtained from FONLL calculations~\cite{theo:Mat09}.}  \label{Fig:2b}
   \end{minipage}
\end{center}
\end{figure}
Figure~\ref{Fig:2b} illustrates the charm and bottom contribution to the single electrons at 0.2, 10 and 14 TeV~\cite{theo:Mat09}. It is remarkable that the crossing point, where bottom starts to dominate over charm seems to be at around the same $\pT$ for all three collision energies. A possible explanation is that at the LHC energies more bottom is produced at small $\pT$ with respect to charm than at RHIC energies, which would bring the crossing point closer to zero. On the other hand the single spectra of bottom and charm are likely to be more similar at the LHC due to the fact that the quark mass is less relevant. This would push the crossing point away from zero. It seems that both mechanisms roughly compensate each other.

ALICE, A Large Ion Collider Experiment, is the dedicated detector to study hot QCD matter in lead-lead collisions at the TeV scale at the CERN-LHC. The initial energy density in the collision zone is expected to be about one order of magnitude higher than at the RHIC facility ($\approx$100~GeV/fm$^3$). The higher energy density allows thermal equilibrium to be reached more quickly and creates a relatively long-lived QGP phase. Therefore, it is expected that most of the in-medium effects will be enhanced. 
The first Pb+Pb data taken in November 2010 at $\snn$~= 2.76 TeV provided first exciting results~\cite{harris}.
Also first heavy-flavor results from $pp$ collisions at $\sqrt{s}$~= 7 TeV have been presented at this workshop~\cite{grelli}.
\begin{figure}[t]
\centering
   \includegraphics[scale=0.403]{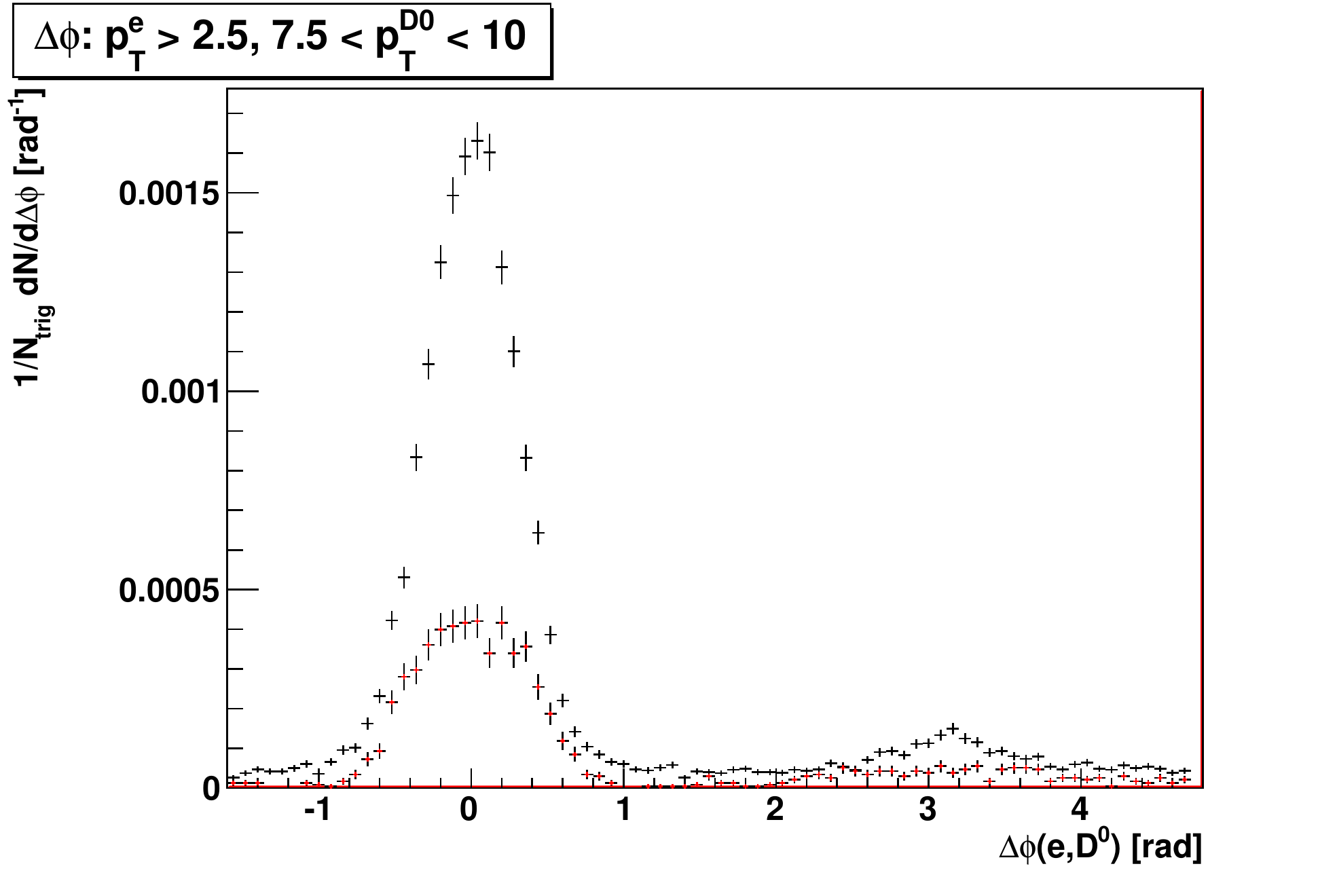}
   \includegraphics[scale=0.39]{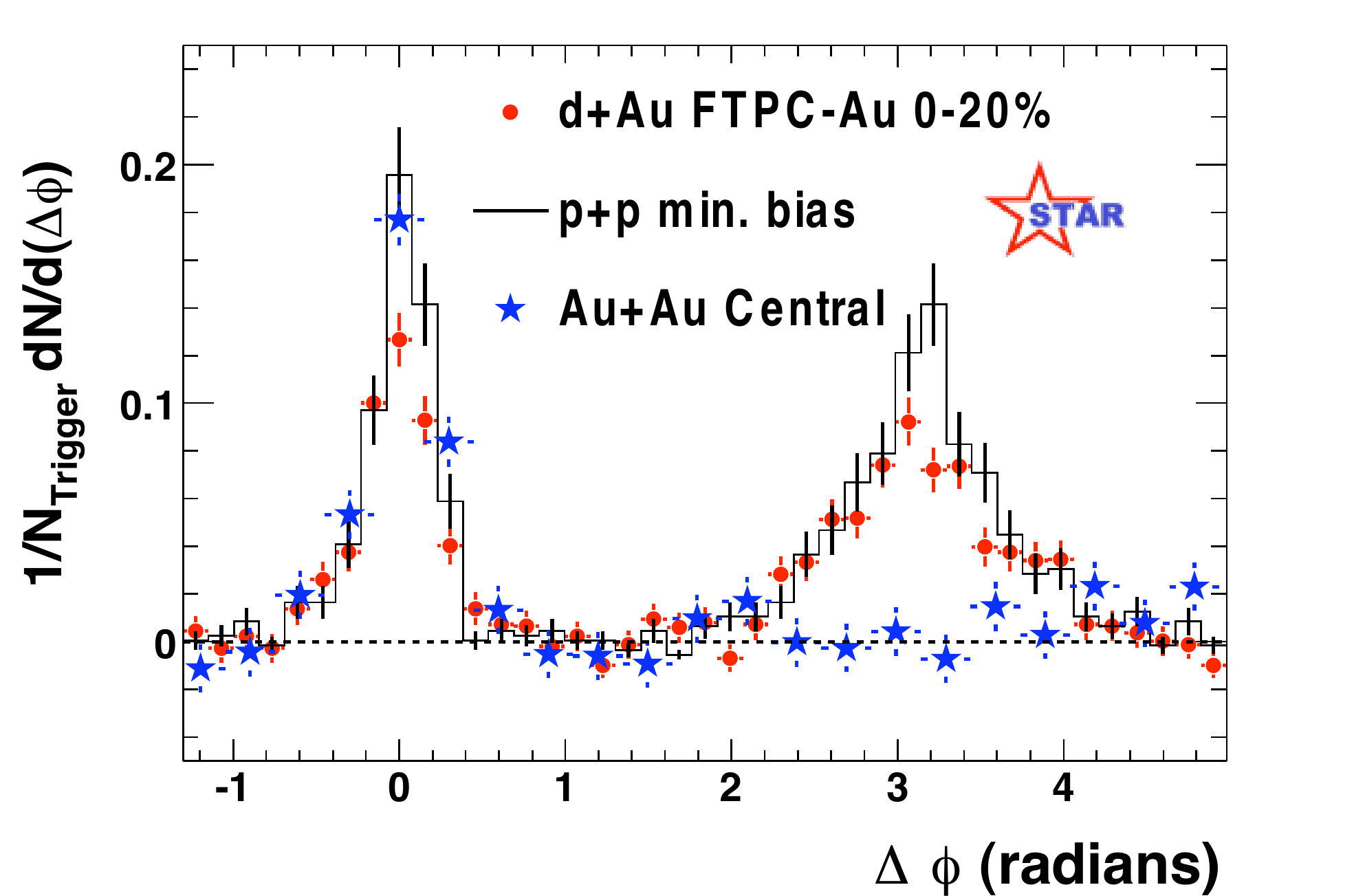}
   \caption{Left panel: Azimuthal angular correlation of single electrons and D$^0$ mesons from PYTHIA simulations of $pp$ interactions (black histogram) and from PYQUEN simulations of the 5\% most central Pb+Pb collisions at $\snn$~= 5.52 TeV (red histogram). 
Right panel: di-hadron azimuthal correlation distribution from $pp$, $d$+Au and Au+Au collisions at $\snn$~= 0.2 TeV measured by the STAR experiment.}  \label{Fig:3}
\end{figure}
The feasibility of heavy-flavor correlations in Pb+Pb collisions at LHC energies were studied using the PYQUEN event generator~\cite{igor}.
The azimuthal angular correlation of single electrons and D$^0$ mesons at mid-rapidity ($|\eta| < 0.9$) is studied in PYTHIA simulations of $pp$ interactions and in PYQUEN simulations of the 5\% most central Pb+Pb collisions at $\snn$~= 5.52 TeV. The left panel of Figure~\ref{Fig:3} shows both correlation distributions.
For the $pp$ collisions the correlation distribution exhibits a pronounced near-side peak, which can be attributed to B decays, and an away-side peak, which arises from charm decays. For Pb+Pb collisions, however, the near-side peak is expected to be suppressed whereas the away-side correlation peak disappear. A similar pattern have be found for di-hadron correlations with the first data from the RHIC facility (Figure~\ref{Fig:3}, right panel)~\cite{star:dAu}, indicating promising results to come from the first LHC data sample.

\begin{figure}[t]
\centering
   \includegraphics[scale=0.42]{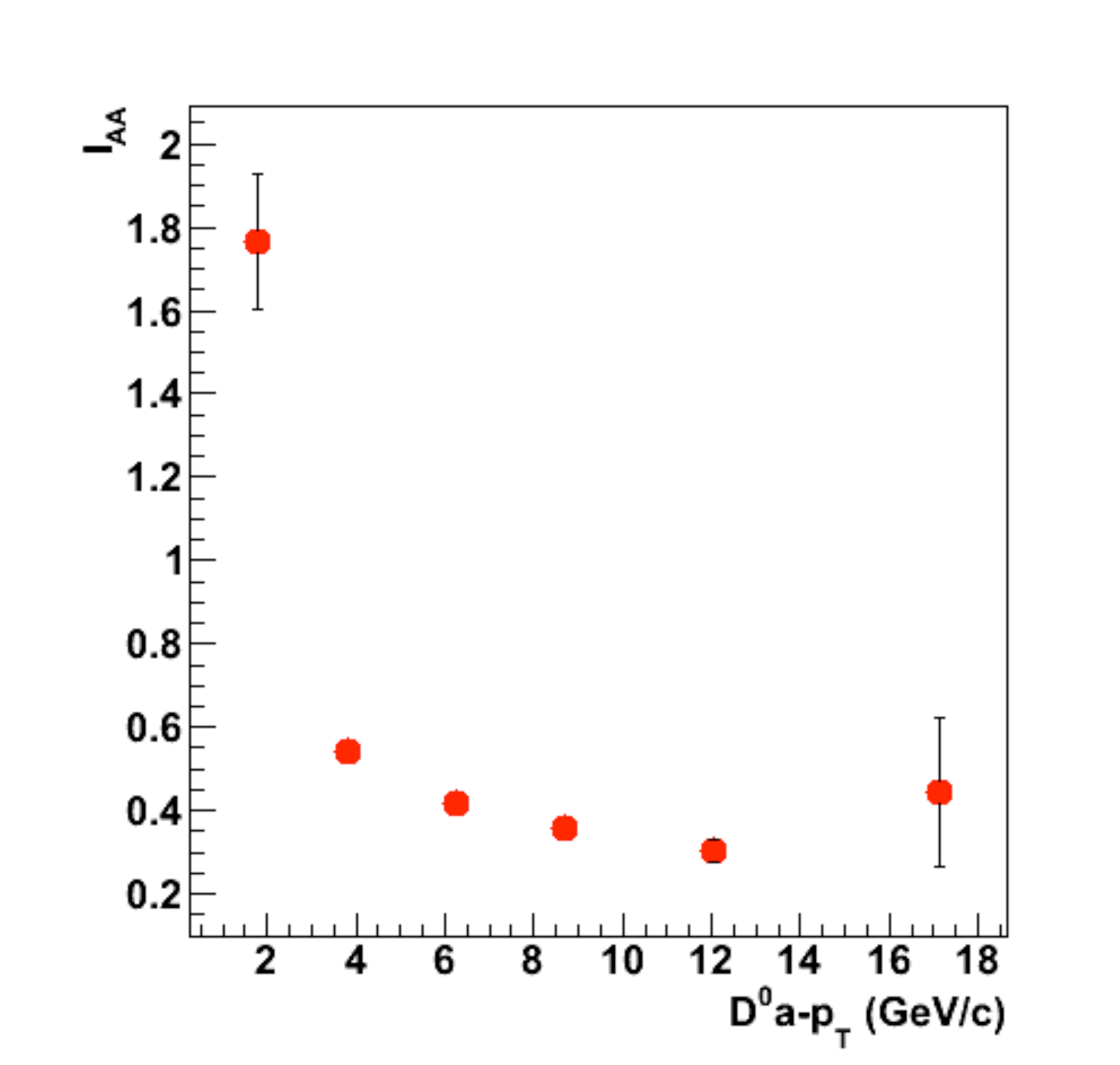}
   \includegraphics[scale=0.623]{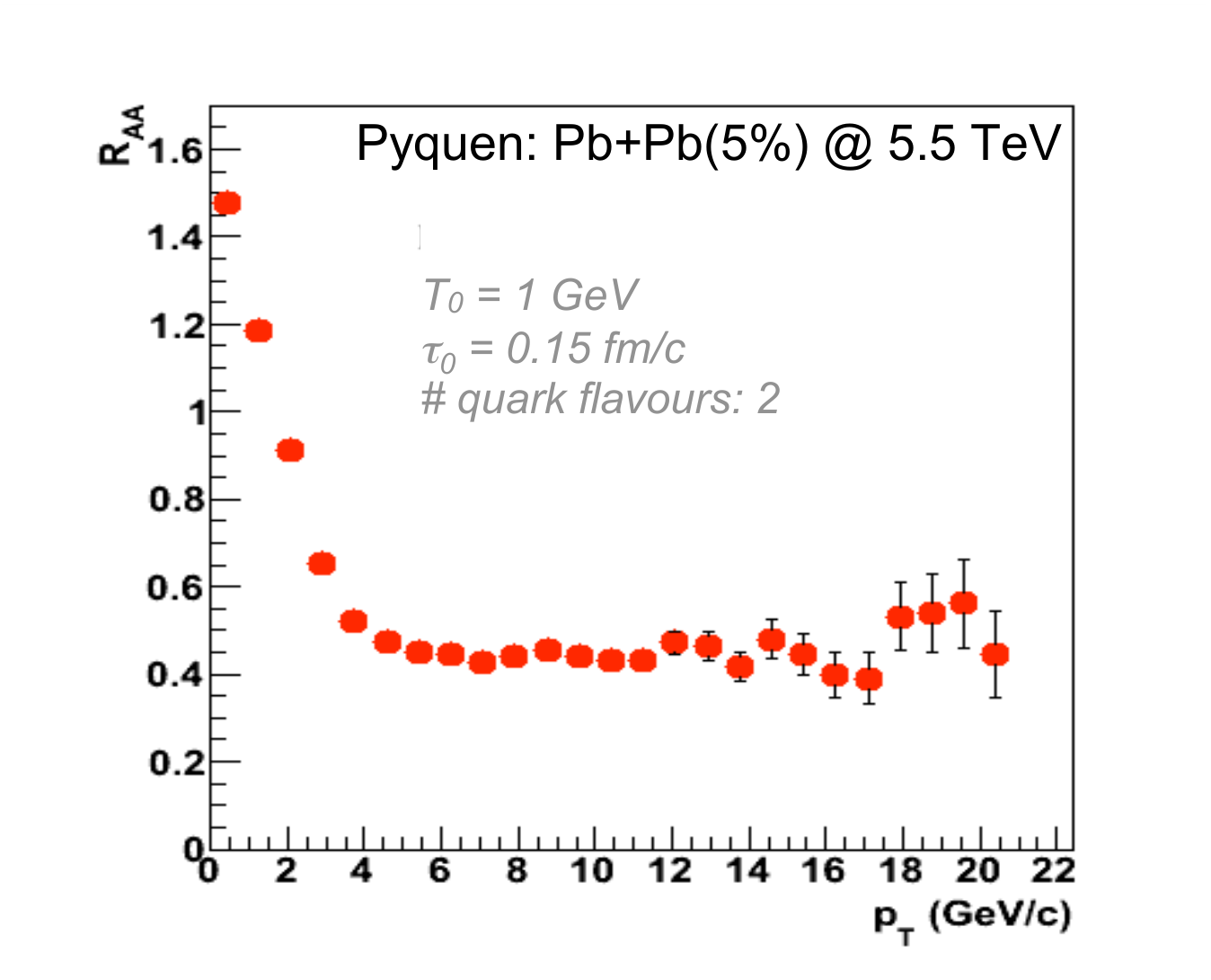}
   \caption{$\iaa$ of the near-side peak (left panel) and $\raa$ of single electrons (right panel) from PYQUEN simulations of the 5\% most central Pb+Pb collisions at $\snn$~= 5.52 TeV. The PYQUEN settings are specified in the right hand panel.}  \label{Fig:4}
\end{figure}

The yield of the near-side peak for an electron trigger $\pT$ of $2 < \pT^{\rm trg-ele} < 4$ GeV/c was extracted from the integral of a Gaussian fit to the data ($\pm 3 \sigma$ range around the peak position). The width of the near-side peak from PYQUEN (Pb+Pb case) is slightly broader compared to PYTHIA ($pp$ case).
The yield as a function of the associated $\pT$ of the D$^0$ mesons is shown in the left panel of Figure~\ref{Fig:4}.
The $\iaa$ decreases and stays constant at around $\pT$~= 8 GeV/c. 
The magnitude of the yield suppression of the near-side peak in central Pb+Pb collisions at $\snn$~= 5.52~TeV is similar to the nuclear modification factor of the single electrons, as illustrated in the right panel of Figure~\ref{Fig:4}. The single electron $\raa$ agrees with the ALICE measurement~\cite{silvia11} and previous theoretical model calculations~\cite{nestor}.
%

\section{Gluon splitting rate}  
In the previous sections is has been shown that the understanding of the heavy-flavor production mechanism is of considerable interest.
Higher order sub-processes such as gluon splitting to $c\bar{c}$ pairs are expected to have a significant contribution to the total open charm yield at LHC energies~\cite{gluonrate}.
The gluon splitting contribution is studied by identifying the {\em charm content in jets}.
By utilizing jet-finder algorithms and isolating the near side of the jet-$D^*$ azimuthal angular correlation, the charm ($D^*$) content in a jet was measured. From the correlation strength on the near side, it is possible to estimate the amount of open charm production which results from gluon splitting. 

\begin{figure}[t]
\centering
   \includegraphics[scale=0.37]{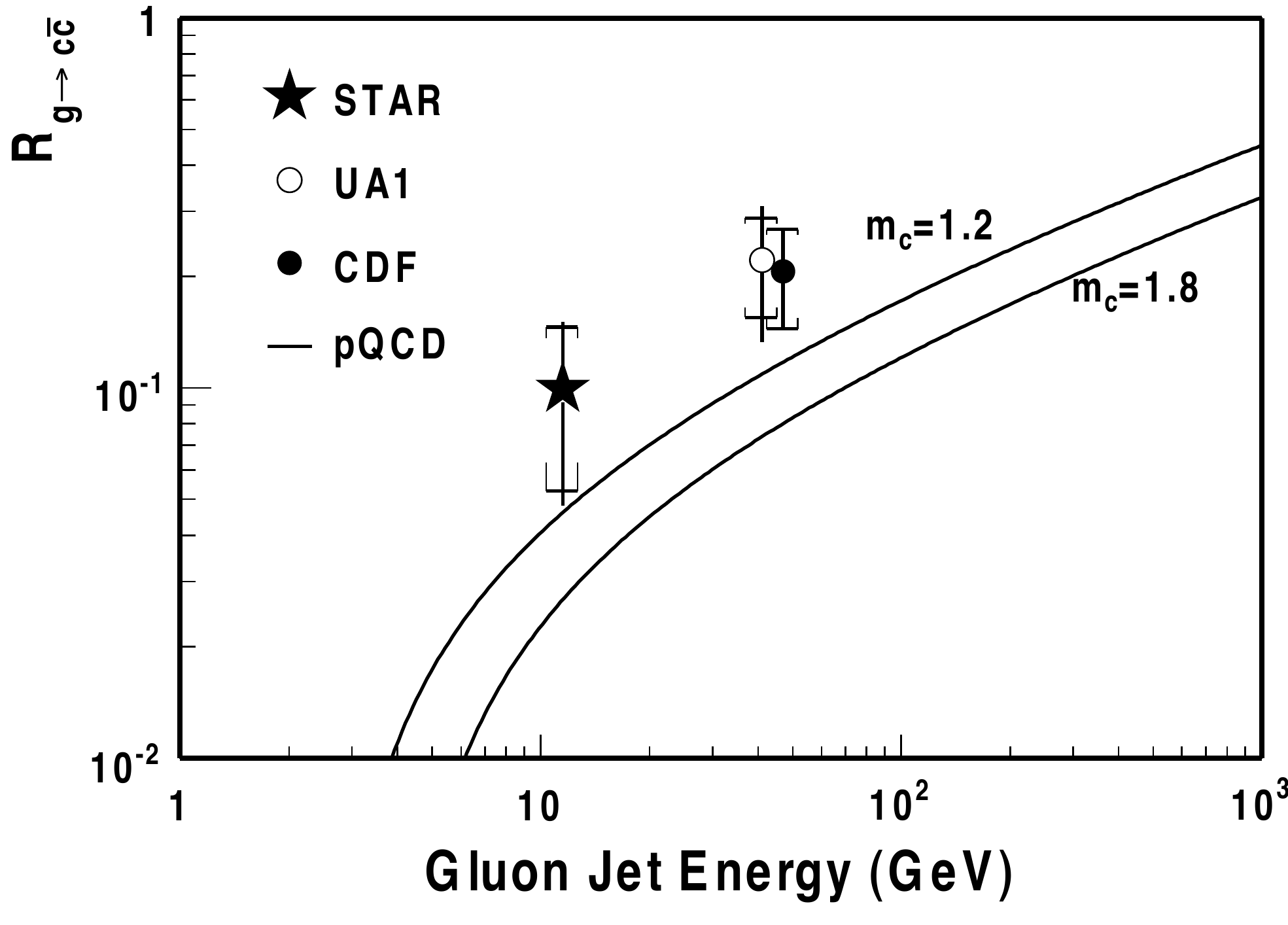}
      \hspace*{4.5mm}
   \includegraphics[scale=0.7]{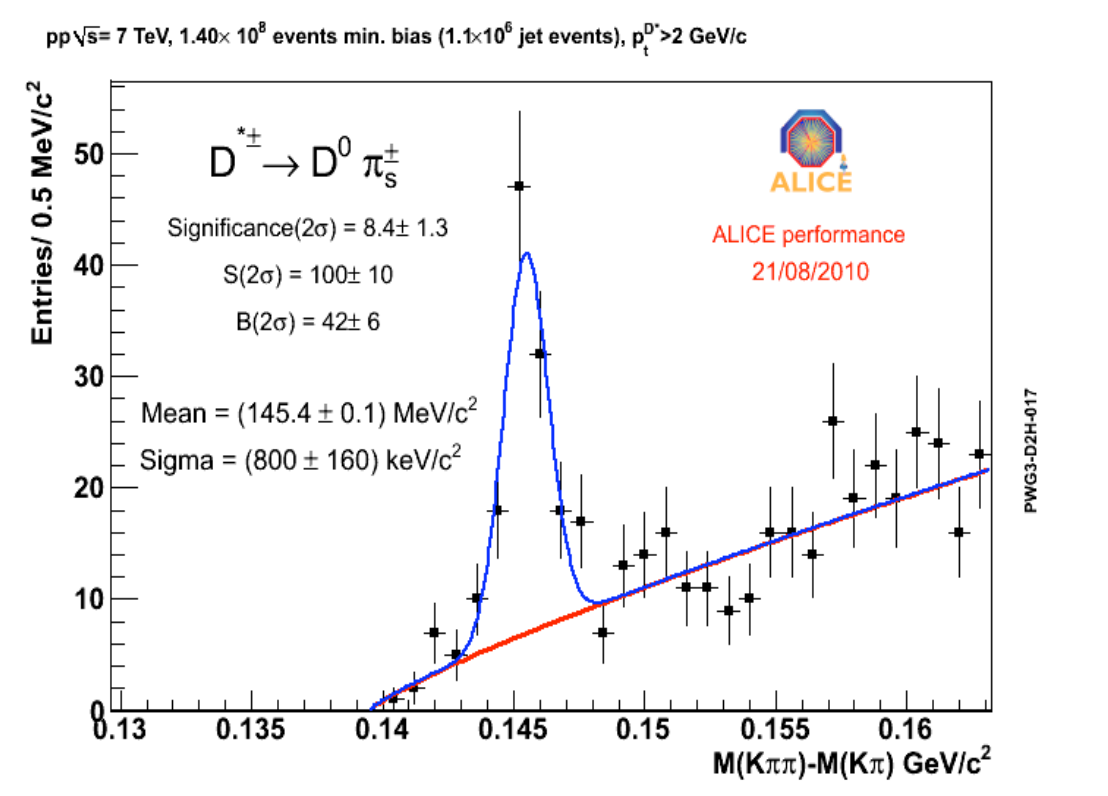}
   \caption{Left: Gluon splitting rate to charm pairs as a function of the gluon jet energy~\cite{star:dstar}, compared with pQCD calculations. Right: Charged $D^*$ signal reconstructed with the ALICE experiment within the cone of jets in $pp$ collisions at $\sqrt{s}$~= 7 TeV~\cite{alice:dstar2}. See text for details.}  \label{Fig:5}
\end{figure}

The LEP experiments~\cite{gs:11, gs:12, gs:13} have studied the D*$^{\pm}$-meson content in jets and demonstrated that the production from Z$^0$ decays in $e^+e^-$ collisions is dominated by D* mesons that carry large fractions of the jet momenta. This finding is consistent with the jets being produced from primary c (anti-)quarks. 
At higher collision energies ($\bar{p}p$ at $\sqrt{s}$~= 1.8 TeV and 630 GeV) the CDF and UA1 experiments have observed D*$^{\pm}$ mesons in the cone of jets with transverse energies larger than 40 GeV~\cite{gs:2, gs:3}. 
Their fractional momenta are found to be smaller, consistent with a different production mechanism in which the D* mesons originate from gluon splitting into $\bar{c}c$ pairs ($g \rightarrow \bar{c}c$ in the initial or final parton shower, with neither of the quarks from the $\bar{c}c$ pair participating in the hard QCD interaction)~\cite{gluonrate}. 
The STAR experiment measured a gluon splitting rate of about 10\% in $pp$ at $\sqrt{s}$~= 200 GeV~\cite{star:dstar}.
The later three measurements are summarized in the left panel of Figure~\ref{Fig:5} compared to perturbative QCD~\cite{theo:glspl1, theo:glspl2, theo:glspl3}.
Moreover, the STAR measurement is consistent with results from MC@NLO calculations~\cite{Misch09}, a QCD computation with a realistic parton shower model.

First performance studies have been conducted at LHC. The right panel of Figure~\ref{Fig:5} shows the charged $D^*$ signal reconstructed within the cone of jets with the ALICE experiment in $pp$ collisions at $\sqrt{s}$~= 7 TeV~\cite{alice:dstar1, alice:dstar2}. This is an important step towards the determination of the charm content in jets at the TeV scale.
%

\section{Summary}  
Heavy quarks are particularly good probes to study the properties of hot QCD matter (especially its transport properties).
RHIC data exhibit a relatively large energy loss of heavy quarks in the medium.
Azimuthal angular correlations of single electron--hadron and single electron--D$^0$ mesons show that bottom decay yields are of comparable magnitude at and above $\approx$5 GeV/c, consistent with FONLL calculations, indicating that in particular bottom is more strongly suppressed than expected.
At the LHC, heavy-flavor particle correlations will allow for the study of the possible modification of the fragmentation function in the medium.
Next-to-leading-order QCD processes, such as gluon splitting, become important at LHC energies and their contribution can be studied through the measurement of the {\em charm content in jets}.
First heavy-ion collisions in November 2010 mark the start of the study of hot quark matter in a new energy domain.

\ack
The author would like to thank the organizers for the nice atmosphere and the stimulating discussions during the workshop.

The European Research Council has provided financial support under the European Community's Seventh Framework Programme (FP7/2007-2013) / ERC grant agreement no 210223.
This work was supported in part by a Vidi grant from the Netherlands Organisation for Scientific Research (project number 680-47-232).


\end{document}